Original Article

# Impact of Hypoglycemia on Glucose Variability Over Time for Individuals with Open Source Automated Insulin Delivery Systems

Arsalan Shahid[1], and Dana M. Lewis[2]

1. CeADAR – Ireland's Centre for AI, University College Dublin, D04 V2N9 Dublin, Ireland
2. OpenAPS, Seattle, WA 98101, USA; Dana@OpenAPS.org

**Arsalan Shahid, PhD, MBA (Corresponding Author)**
Technology Solutions Lead at CeADAR
Ireland's Centre for AI at University College Dublin
D04 V2N9 Dublin, Ireland
Email: arsalan.shahid@ucd.ie
ORCID: 0000-0002-3748-6361

**Dana M. Lewis, BA**
Independent Researcher
#OpenAPS
Seattle, WA, USA
Email: Dana@OpenAPS.org
ORCID: 0000-0001-9176-6308

**Conflicts of Interest:** The authors declare no financial conflict of interest. DML is a volunteer developer of one of the open-source AID systems, OpenAPS. The funders had no role in the design of the study; in the collection, analyses, or interpretation of data; in the writing of the manuscript; or in the decision to publish the results.

**Author Contributions:** Conceptualisations, AS, DML; methodology; AS, DML; software, AS, DML; validation; AS, DML; formal analysis; AS, DML; investigation; AS, DML; resources; AS, DML; data curation, AS, DML; writing-original draft preparation, AS, DML; writing - review and editing, AS, DML; visualization, AS; project administration, DML; funding acquisition, AS, DML. All authors have read and agreed to the published version of the manuscript.

**Institutional Review Board Statement:** Ethics approval was not required as this work evaluated data from a retrospective, already collected and anonymized dataset.

**Data Availability Statement:** All programming scripts and tools developed for the analysis in this paper are made public and online, with each source cited within the paper. Data accessed for this paper is part of the OpenAPS Data Commons.

**Funding:** This work was in part performed under a grant from The Leona M. and Harry B. Helmsley Charitable Trust.

**Acknowledgments:** Thank you to members of the diabetes community who have donated their data, from a variety of open source automated insulin delivery (AID) systems, to the OpenAPS Data Commons.

**Figures and Tables:** 3 figures (including individual panels, 6 figures total)

# Abstract

**Background:** This study investigates glucose conditions preceding and following various hypoglycemia levels in individuals with type 1 diabetes using open-source automated insulin delivery (AID) systems. It also seeks to evaluate relationships between hypoglycemia and subsequent glycemic variability.

**Methods:** Analysis of continuous glucose monitor (CGM) data from 122 individuals with type 1 diabetes using open-source AID from the OpenAPS Data Commons was conducted. The study comprehensively analyzed the effects of hypoglycemia on glycemic variability, covering time periods before and after hypoglycemia.

**Results:** Glucose variability normalization post-hypoglycemia can take up to 48 hours, with severe hypoglycemia (40-50 mg/dL) linked to prolonged normalization. A cyclical pattern was observed where hypoglycemia predisposes individuals to further hypoglycemia, even with AID system use. A rise in glucose levels often precedes hypoglycemia, followed by an elevated mean time above range (TAR) post-severe hypoglycemia, indicating a 'rebound' effect. Different hypoglycemia categorization approaches did not show significant differences in glycemic variability outcomes. The level of hypoglycemia does influence the pattern of subsequent glucose fluctuations.

**Conclusion:** Hypoglycemia, especially at lower levels, significantly impact subsequent glycemic variability, even with use of AID systems. This should be studied further with a broader set of commercial AID systems to understand if these patterns are true of all types of AID systems. If these patterns occur in all types of AID systems, it underscores potential opportunities for enhancements in AID algorithms and highlights the importance of educating healthcare providers and people with diabetes about post-hypoglycemia glucose variability.

# Introduction

In 2021, approximately 8.4 million people were living with type 1 diabetes, a figure projected to nearly double by 2040[1]. Managing type 1 diabetes (T1D) involves continuous glucose monitoring and insulin administration, either manually or through continuous subcutaneous insulin infusion (CSII) pumps. Recent advancements like continuous glucose monitors (CGM) and automated insulin delivery (AID) systems are enhancing glycemic outcomes and quality of life for those with diabetes[2]. However, the delay in insulin's peak effect and the pharmacokinetic curve[3] present challenges in synchronizing dosage with variables such as food intake and physical activity[4], which can influence glucose-related fluctuations and result in hypoglycemia.

Over 25 glucose variability (GV) and glucose-related metrics are prevalent in diabetes research. These metrics, assessable via open-source tools like cgmquantify and CGM-GUIDE, help to evaluate glucose data and understand glycemic variability[5–14]. AID systems improve time in target glucose range (TIR) as well as time above range (TAR) and time below range (TBR); however, hypoglycemia (time spent below range) still occurs for a variety of reasons[3]. The American Diabetes Association categorizes hypoglycemia into "Level 1" (<70 mg/dL but >54 mg/dL) and "Level 2" (between 40 and <54 mg/dL)[15]. Understanding the correlation of hypoglycemia with glucose variability is a possible focus area for improving clinical outcomes and quality of life, as previous studies indicate a predictive relationship between glucose variability and severe hypoglycemia[16, 17].

Using an open-source AID dataset[18] this study explores the interplay between hypoglycemia, hyperglycemia, and glycemic variability-related metrics of continuous glucose monitor (CGM) sensor data. This includes ADA's Level 1 and Level 2 designations as well as additional levels of 40-50 mg/dL; 50-60 mg/dL; and 60-70 mg/dL to evaluate whether there are distinct pattern differences associated with different levels of hypoglycemia. We sought to evaluate glycemic variability and identify patterns in CGM sensor glucose levels before and after different levels of hypoglycemia in users of open-source AID systems in real-world data. We hypothesize that an understanding of patterns in glucose levels and GV-related metrics before and after hypoglycemia could provide potential insights for AID system enhancement, specifically related to improving post-hypoglycemic glycemic variability and reducing the occurrence of additional hypoglycemia in individuals with insulin-requiring diabetes.

# Methods

## Dataset collection and description

The primary dataset used in this paper originates from the OpenAPS Data Commons[19] which allows for anonymized data donation[20] from real-world users of open-source AID (OS-AID) systems (which include OpenAPS, AndroidAAPS/AAPS, and Loop). Further details of the data collection, cleaning process, and more detailed characterization of this version (n=122 users) of the dataset can be found in our previously published papers[21, 22], and all analysis described below uses open-source processing scripts[23].

**Glucose analysis metrics and statistical tests**

The statistical and variability metrics for glucose analysis calculated in this paper include mean, minimum, and maximum of CGM sensor data, the first, second, and third quartiles, and the interquartile range. We also measure the interday standard deviation (SD) along with time outside range for both hypoglycemia (TOR<70) and hyperglycemia (TOR>180), and time in range (TIR). Moreover, the analysis considers the J_index (which stresses both the importance of the mean level and variability of glycemic levels)[24] as well as LBGI, HBGI, Coefficient of Variation (CV) and Glucose Management Indicator (GMI) [25] to evaluate glycemic variability.

To probe the relationship between hypoglycemia and glucose variability, we utilized a series of statistical tests, including the Shapiro-Wilk (SW) test for testing the normality of a data distribution; the Z-test which compares the means of hypoglycemia and glucose variability, to investigate any significant differences; and the Kolmogorov-Smirnov (KS) test, a nonparametric test to determine if two datasets differ significantly.

**Experimental workflow**

A three-phased structured experimental workflow (*Figure A1, Appendix*) has been designed to investigate the impact of hypoglycemia, at different levels, on glucose variability (GV).

- Phase 1 — *Design Decisions:* We selected GV metrics — TIR, TOR<70, TOR>180, LBGI, HBGI, SD, POR, J_Index, CV, and GMI. We defined hypoglycemia levels at five different ranges: 40-50 mg/dL, 50-60 mg/dL, 60-70 mg/dL, 40-54 mg/dL, and 55-70 mg/dL. This includes more granular, 10-point ranges, as well as ADA-defined level 1 and 2 ranges, to determine whether a tighter hypoglycemia range influenced resulting patterns.
- Phase 2 — *Data Preparation:* We used identical pre-processing and preparation methods from previous work[22].
- Phase 3 — *Out-of-Whack Analyser* (visualized in Figure 1): We computed GV-related metrics before and after each instance of hypoglycemia at various intervals: 3, 6, 12, 24, and 48 hours. A hypoglycemic event, or instance of hypoglycemia, was defined as any instance where sensor glucose fell below or into the specified range. Successive data points of hypoglycemia after the first were also identified, and the entire contiguous sequence is considered a hypoglycemia segment, with the GV calculated before and after this segment. In some cases, where sensor glucose returned above range then returned into the defined hypoglycemia range, a second hypoglycemic segment was recorded and separately analyzed. This provided a comprehensive view of the impact of different levels of hypoglycemia on GV over different timeframes. Subsequently, we calculated mean statistics for each individual, providing a representative measure for comparison across subjects. Lastly, we conducted a distribution analysis to visualize and interpret the mean GV distributions before and after hypoglycemia.

**FIGURE 1:** Process flow diagram for three-stage out-of-whack analyser

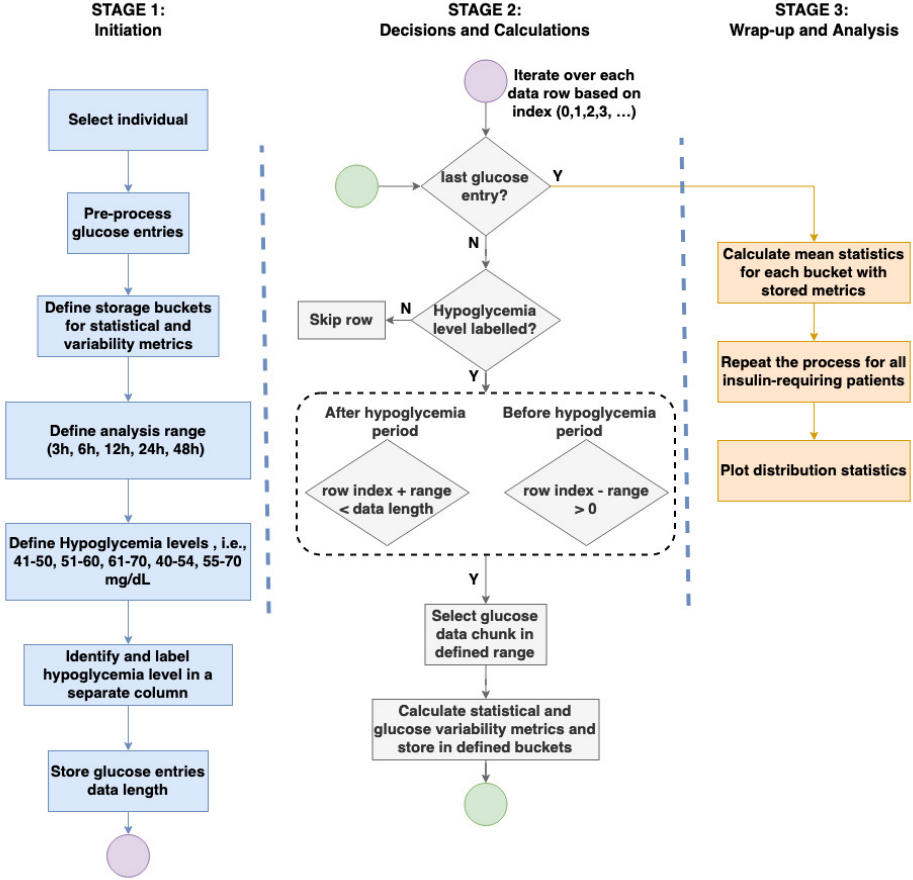

Data is organized into tables (*Appendix*) to provide a comprehensive view of glucose analysis and variability metrics including TBR, TIR, TOR, SD, and J_Index for various time periods surrounding the instances of hypoglycemia, respectively. To quantify these metrics, we incorporated several statistical measures: mean, standard deviation (SD), minimum (Min), first quartile (25%), median (50%), third quartile (75%), maximum (Max), Shapiro-Wilk test (SW-test), Z-test, and Kolmogorov-Smirnov test (KS-test).

The organization of data spans from 48 hours before (-48 hours) to 48 hours after (+48 hours) hypoglycemia, with interval granularity increasing as the event approaches. These periods are further subdivided into five categories of hypoglycemia: 40-50 mg/dL, 50-60 mg/dL, 60-70 mg/dL, 40-54 mg/dL, and 55-70 mg/dL.

To facilitate visual inspection and trend analysis, figures are designated to represent these data across the first three hypoglycemia categories (40-50 mg/dL, 50-60 mg/dL, and 60-70 mg/dL), including Figures 2a, 2b, 2c, 3a, 3b (and *Appendix A2b and A2a*) for TBR, TIR, TAR, SD, J_Index, HBGI, and LBGI, respectively. Additional figures (*Appendix: A3a, A3b, A3c, A4a, A4b, A5a, and A5b*), focus on the remaining hypoglycemia levels (40-54 mg/dL and 55-70 mg/dL).

## Results

This analysis of instances of hypoglycemia in the OpenAPS Data Commons dataset revealed how different levels of hypoglycemia affect glycemic variability over time. The levels of hypoglycemia have an impact on the patterns of glucose levels both before and after such events, with notable variations in GV metrics across different categories or levels of hypoglycemia. Specifically, the data shows that glucose levels often begin to fluctuate significantly as hypoglycemia approaches, with this variability extending up to 48 hours post-event before normalizing. These patterns do not follow a normal distribution and there are statistically significant differences in glucose metrics between the various hypoglycemia severity categories. Although we calculated CV and GMI, we found no clear pattern connecting them to hypoglycemia at the studied timeframes.

Notably, the level of hypoglycemia has a significant impact on the patterns of glucose levels before and after hypoglycemia (*Appendix Table A1*). In the periods leading up to the hypoglycemia (indicated by the negative time intervals), the average TBR increases as the time period approaches the event (Figures 2A and *A3a*). In other words, a prolonged period of lower glucose levels often precedes other instances of hypoglycemia, regardless of the level of hypoglycemia.

**FIGURE 2:** Comparison of distributions (box plots) for percentage of A) time below range (TBR or TOR<70), B) time in range (TIR), and C) time above range (TAR > 180 mg/dL), before and after hypoglycemia at levels of 40-50 mg/dL, 50-60 mg/dL, and 60-70 mg/dL.

*A) Distributions for percentage of Time Below Range (TBR or TOR<70) before and after hypoglycemia.*

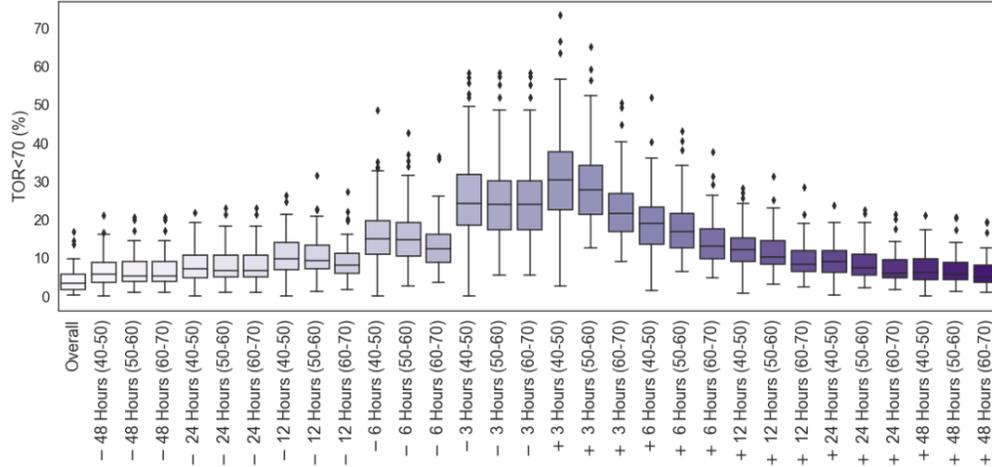

*B) Distributions for percentage of time in range (TIR) before and after hypoglycemia.*

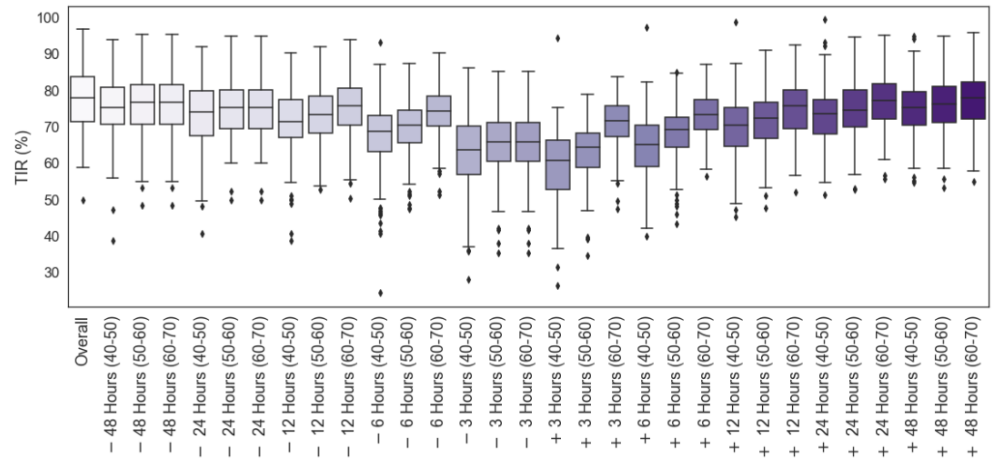

*C) Distributions for percentage of time above range (TAR or TOR>180) before and after hypoglycemia.*

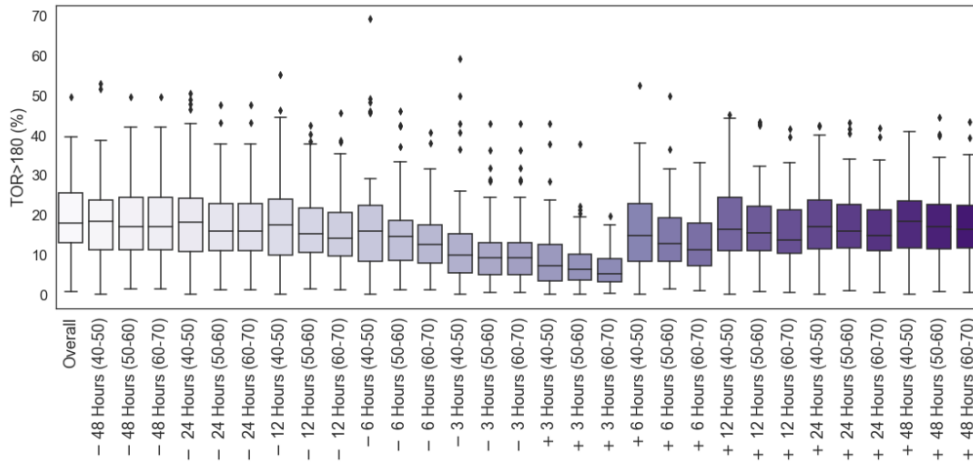

However, the actual mean values for TBR differ between the categories. Lower hypoglycemia (40-50 mg/dL) are generally preceded by longer periods of low glucose levels than less severe lows (60-70 mg/dL). This indicates that the level of hypoglycemia may be influenced by the duration and severity of the preceding low-glucose period.

In the periods following the hypoglycemia (indicated by the positive time intervals), the average TBR decreases, indicating a recovery in glucose levels. This recovery pattern is observed across all severity categories, suggesting that glucose levels tend to stabilize after hypoglycemia, regardless of the event's severity. The speed and extent of this recovery appear to vary between the categories. Following more severe hypoglycemia (40-50 mg/dL), the average TBR remains higher in the initial hours after the event, suggesting a slower recovery. In contrast, following less severe events (60-70 mg/dL), the average TBR decreases more rapidly, indicating a quicker recovery.

However, it can be observed from the box-plot distributions in Figures 2A and *A3a* that it may take up to 48 hours to completely stabilize TBR as per normal (or overall) TBR distribution.

**FIGURE 3**: Comparison of distributions (box plots) for A) standard deviation (SD) and B) J_Index, before and after hypoglycemia at levels of 40-50 mg/dL, 50-60 mg/dL, and 60-70 mg/dL.

A) *Distributions for standard deviation (SD) before and after hypoglycemia.*

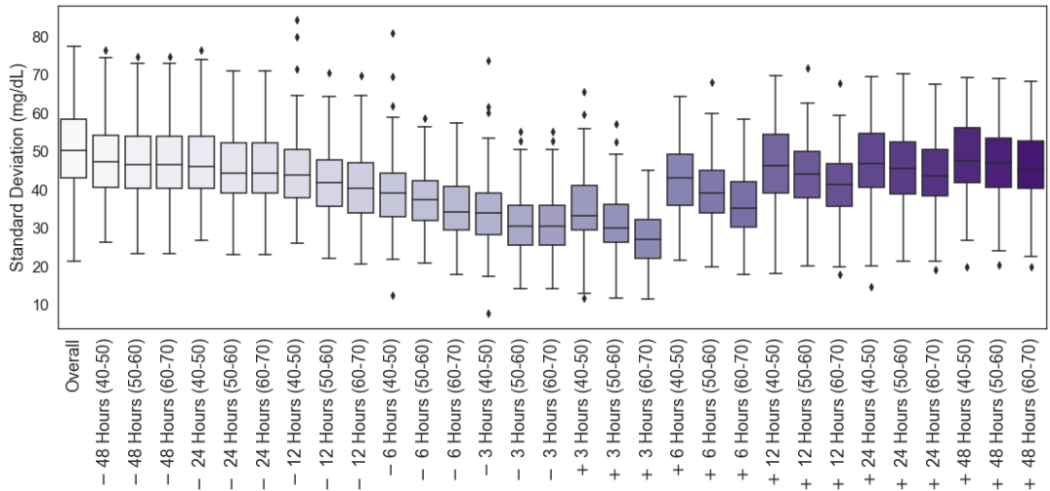

B) *Distributions for J_Index before and after hypoglycemia*

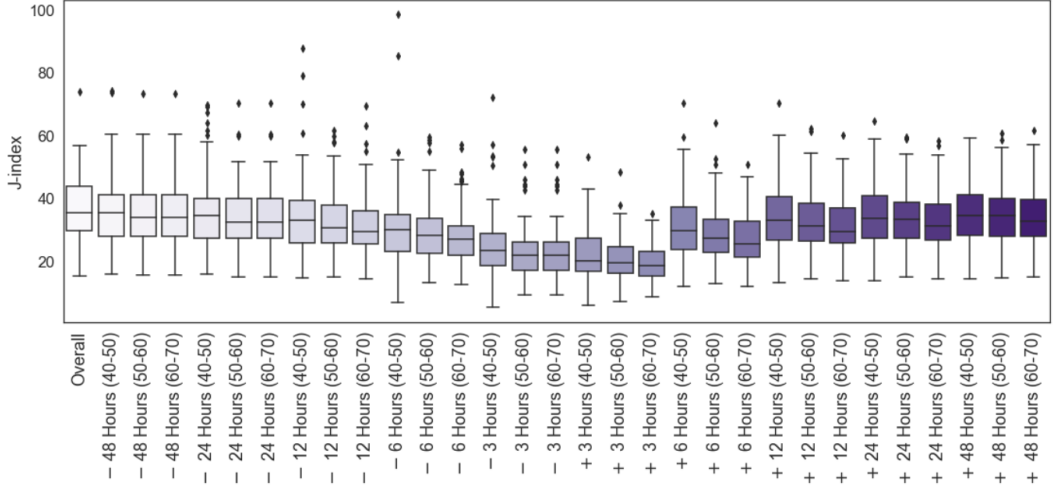

When looking at hypoglycemia 40-50 mg/dL, we observe that the mean TIR tends to decrease with shorter time intervals around the hypoglycemia (*Appendix Table A2*). For example, the mean TIR is 74.89% for the 48-hour window but decreases to 62.25% for the 3-hour window. This suggests that lower hypoglycemia have a substantial impact on TIR, particularly in the immediate hours surrounding the event. For less severe hypoglycemia (50-60 mg/dL and 60-70 mg/dL), a similar pattern can be observed, with TIR decreasing as the time interval around the hypoglycemia becomes shorter. The SW-test, KS-test, and Z-test show statistically significant differences, which suggests that the TIR distribution varies significantly depending on the severity of the hypoglycemia and the time interval around the event.

It can also be observed from the 48-hours before the hypoglycemia that individuals tend to experience higher glucose levels in the days leading up to a hypoglycemia, irrespective of the event's severity (*Appendix Table A3*). In the period following the hypoglycemia (from +3 hours to +48 hours), the mean TAR first decreases and then increases across all categories. The severity of the hypoglycemia appears to influence these patterns. Following hypoglycemia in the lowest range (40-50 mg/dL), the mean TAR tends to be higher, indicating a greater tendency towards hyperglycemia, and what colloquially people with diabetes might describe as a "rebound". Following less severe hypoglycemia (60-70 mg/dL), the mean TAR>180 tends to be lower, indicating a lesser tendency towards rebound hyperglycemia.

In terms of standard deviation, as hypoglycemia approaches, the SD decreases, suggesting that glucose levels become less variable (*Appendix Table A4*). For instance, the mean SD decreases from 48.04 mg/dL at -48 hours to 34.43 mg/dL at -3 hours for the 40-50 mg/dL category. This pattern is observed across all categories of hypoglycemia. Lower glucose levels are associated with smaller SD values both before and after the event.

The mean J_Index score tends to decrease as the time period gets closer to the hypoglycemia, both before and after the event (*Appendix Table A5*). This further confirms that the glucose outcomes are less optimal closer to the hypoglycemia. There is also a general trend of decreasing J_Index score associated with lower glucose, indicating worse glucose outcomes associated with increased severity of hypoglycemia.

## Discussion

Hypoglycemia remains a significant challenge in diabetes management, and understanding the patterns and after-effects of different levels of hypoglycemia is critical to improving therapeutic approaches and improving outcomes. In our study evaluating the relationship between different levels of hypoglycemia and resulting patterns in glucose outcomes using CGM data from PWD using OS-AID systems, we observed that glucose levels can take up to 48 hours to stabilize post-hypoglycemia. More pronounced lows (e.g., 40-50 mg/dL) correlate with prolonged normalization times. This observation extends the common understanding that the after-effects of hypoglycemia are not immediate, and the 'shadow' of an episode of lower glucose can last for up to two days.

Furthermore, our data underscore the cyclic nature of hypoglycemia: an episode of low glucose is observed to be associated with subsequent hypoglycemia. This trend persists even with advanced diabetes therapies like automated insulin delivery (AID) systems. Particularly at levels of 40-50 mg/dL, in this dataset we observed a greater risk of recurring hypoglycemia within 48 hours than at slightly higher levels (60-70 mg/dL). A similarly noteworthy trend is also the heightened glucose levels observed in the days leading up to hypoglycemia. As hypoglycemia draws near, the duration spent above the optimal range decreases. Following severe hypoglycemia (40-50 mg/dL), we observed an elevated mean time above range (TAR), a phenomenon often termed a "rebound" in the diabetes community. This mirrors previous studies and understandings of hypoglycemia in T1D where higher fluctuations (thus increased GV) increase the likelihood of subsequent instances of hypoglycemia; yet it remains unclear whether this is a cause or a consequence itself of hypoglycemia[26].

The distribution of time in range (TIR) is influenced by the severity of the preceding hypoglycemia. Higher levels of hypoglycemia disturb TIR to a lesser degree than lower hypoglycemia, potentially due to their closer proximity to the optimal range (70-180 mg/dL). While ADA defines hypoglycemia based on two levels, with "Level 1" being <70 mg/dL and >54 mg/dL and "Level 2" <54 mg/dL, we sought to explore the relationship in more narrow categories and evaluated glycemic variability and outcomes related to hypoglycemia at ranges of 40-50 mg/dL, 50-60 mg/dL, and 60-70 mg/dL. This categorization was in pursuit of a deeper understanding of how the intensity of lows impacts glycemic variability and improves the quantification of hypoglycemia associated with glycemic variability, which has not previously been analyzed.

While the relationship between hypoglycemia and glycemic variability is increasingly studied using CGM data, glycemic variability is often assessed for an overarching period (e.g. days to weeks or longer)[27] and/or blocks of time (e.g. day versus night), rather than specifically in relationship to instances of hypoglycemia[28]. Previous literature investigating the glycemic variability and hypoglycemia relationship often relies on computations from fingerstick blood glucose testing (SMBG)[16], although there are a few studies increasingly using CGM data[26], more closely mirroring our analyses albeit in a much smaller dataset. We found that the level of hypoglycemia does indeed affect the intensity of glycemic variability distributions, but we found no difference in the outcomes between the three, smaller ranges (40-50, 50-60, 60-70 mg/dL) and the broader Level 1 (<70, >54 mg/dL) and Level 2 (<54 mg/dL) categorization historically used. If these patterns are repeatedly observed (as planned in future studies) across other AID datasets, then future glycemic variability analyses in relation to hypoglycemia likely can use the more general Level 1 and Level 2 hypoglycemia categorizations. It is also worth highlighting that clinical readers may have a perception of hypoglycemia event categorizations related to historical definitions of hypoglycemia, validated by fingerstick blood glucose testing (often referred to as SMBG). In our study, we found that evaluating instances of hypoglycemia separated by a single data point (e.g. where sensor glucose returned above 70 mg/dL for a single data point then dipped <70 mg/dL again), rather than defining this time period collectively as a single instance of hypoglycemia, did not influence the patterns observed, especially as we focused on the broader time periods of 3 hours or more following the segment(s) of hypoglycemia in sensor glucose levels.

This study provides insights into the patterns of glycemic variability among insulin-requiring individuals using OS-AID systems across different levels of hypoglycemia. Future work is underway, and as-of-yet unpublished data confirms that these patterns also exist among users of multiple types of commercial AID systems. Additional analysis in the future will also evaluate potential determinants of these patterns, including assessment of insulin activity prior to hypoglycemia as well as patterns of carbohydrate intake related to treating hypoglycemia. These studies could possibly then collectively contribute to the development of strategies that minimize these patterns, focusing on behavioral responses to hypoglycemia, insulin activity levels preceding and following hypoglycemia, and both user and system interventions that might perpetuate recurrent hypoglycemia.

**Limitations**

This work leverages retrospective data from insulin-requiring individuals using open-source automated insulin delivery (AID) systems. Further, the analysis to date focused primarily on establishing patterns related to glucose data, due to the complex, retrospective nature of the real-world dataset where variables such as IOB are not perfectly logged due to the intermittent upload nature of the dataset. As such, forthcoming work will replicate using different datasets that include commercial AID user data (already underway) to determine whether these glycemic variability patterns related to hypoglycemia also pertain to non-AID users.. Target glucose level, which can be customized by users of OS-AID to different levels than is available in commercial AID systems, was not factored into the current analysis; but the user-set target of OS-AID or targets of commercial AID systems should be considered in future analyses related to patterns of hypoglycemia.

For this work, we chose a 48-hour window preceding and following hypoglycemia, with a resolution down to 3 hours pre- and post-hypoglycemia. In some cases, there appears to be an increase in metrics around the 6-hour mark, such as TBR <70 mg/dL and TAR >180 mg/dL. Future work should evaluate other time intervals at smaller time intervals to understand whether this increase is an artifact of the chosen time intervals or a significant increase at that particular time. This may also be influenced by the time of insulin action, which is often around 5-7 hours when used in AID systems and is another factor to be explored in future work.

**Conclusions**

This paper investigated the dynamics of different levels of hypoglycemia and glycemic variability in people with Type 1 diabetes using open-source automated insulin delivery (AID) systems. Our findings revealed that a single instance of hypoglycemia is related to subsequent hypoglycemia in a short timeframe and that hyperglycemia often precedes significant hypoglycemia. This study showed that glucose variability can take up to 48 hours to stabilize post-hypoglycemia, with more pronounced hypoglycemia correlating with prolonged normalization times, even with use of AID systems. Such insights suggest that future refinements in automated insulin delivery (AID) system algorithms could enhance glucose management following instances of hypoglycemia and periods of more extreme glucose fluctuations, and that there are areas of opportunity for increasing education for diabetes care providers and people living with diabetes about post-hypoglycemia timeframes.

1    SUPPLEMENTARY APPENDIX
2
3    **FIGURE A1:** Experimental Workflow

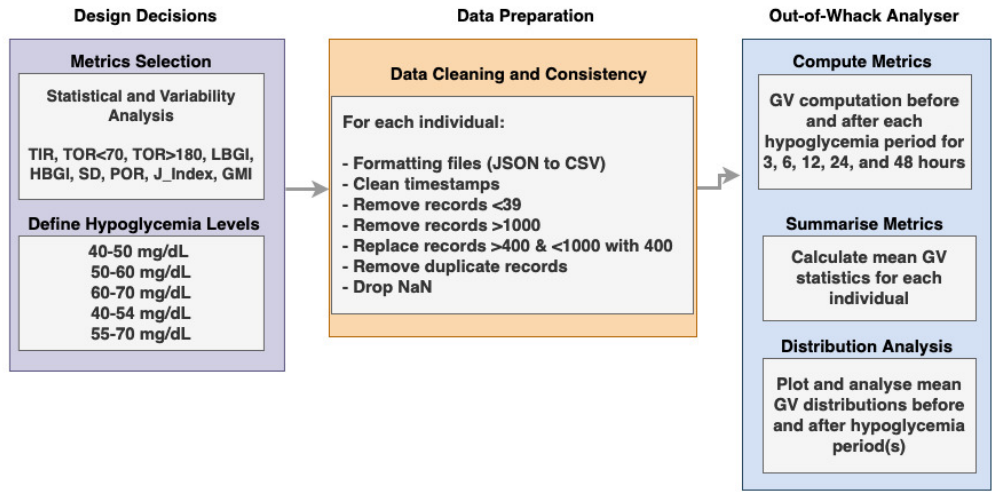

4
5

Table A1: **Statistics of glucose analysis and variability metrics for Time Below Range (TBR) or Time Outside Range less than 70 mg/dL (TOR<70%) for various time periods surrounding the instances of hypoglycemia, respectively.**

| | Mean (%) | SD (%) | Min (%) | 25% | 50% | 75% | Max (%) | SW-test | Z-test | KS-test |
|---|---|---|---|---|---|---|---|---|---|---|
| Overall | 4.09 | 3.06 | 0.23 | 1.74 | 3.33 | 5.69 | 16.97 | <0.05 | | |
| - 48 Hours (40-50) | 6.48 | 3.79 | 0 | 3.65 | 5.82 | 8.81 | 21.18 | <0.05 | <0.05 | <0.05 |
| - 48 Hours (50-60) | 6.44 | 3.82 | 0.91 | 3.82 | 5.29 | 8.94 | 20.68 | <0.05 | <0.05 | <0.05 |
| - 48 Hours (60-70) | 6.44 | 3.82 | 0.91 | 3.82 | 5.29 | 8.94 | 20.68 | <0.05 | <0.05 | <0.05 |
| - 24 Hours (40-50) | 8.02 | 4.08 | 0 | 4.92 | 7.13 | 10.74 | 21.89 | <0.05 | <0.05 | <0.05 |
| - 24 Hours (50-60) | 7.99 | 4.2 | 1.07 | 5.12 | 6.77 | 10.63 | 22.91 | <0.05 | <0.05 | <0.05 |
| - 24 Hours (60-70) | 7.99 | 4.2 | 1.07 | 5.12 | 6.77 | 10.63 | 22.91 | <0.05 | <0.05 | <0.05 |
| - 12 Hours (40-50) | 10.58 | 5.1 | 0 | 6.91 | 9.76 | 13.92 | 26.39 | <0.05 | <0.05 | <0.05 |
| - 12 Hours (50-60) | 10.46 | 5.06 | 1.36 | 7.1 | 9.2 | 13.25 | 31.56 | <0.05 | <0.05 | <0.05 |
| - 12 Hours (60-70) | 9.07 | 4.33 | 1.84 | 5.99 | 8.07 | 11.19 | 27.19 | <0.05 | <0.05 | <0.05 |
| - 6 Hours (40-50) | 16.23 | 7.96 | 0 | 10.95 | 14.87 | 19.69 | 48.61 | <0.05 | <0.05 | <0.05 |
| - 6 Hours (50-60) | 15.72 | 7.31 | 2.72 | 10.55 | 14.67 | 19.28 | 42.67 | <0.05 | <0.05 | <0.05 |
| - 6 Hours (60-70) | 13.26 | 5.98 | 3.68 | 8.91 | 12.36 | 16.21 | 36.45 | <0.05 | <0.05 | <0.05 |
| - 3 Hours (40-50) | 25.93 | 11.85 | 0 | 18.58 | 24.15 | 31.8 | 58.33 | <0.05 | <0.05 | <0.05 |
| - 3 Hours (50-60) | 25.02 | 10.71 | 5.45 | 17.44 | 23.98 | 30.03 | 58.24 | <0.05 | <0.05 | <0.05 |
| - 3 Hours (60-70) | 25.02 | 10.71 | 5.45 | 17.44 | 23.98 | 30.03 | 58.24 | <0.05 | <0.05 | <0.05 |
| + 3 Hours (40-50) | 31.84 | 11.92 | 2.78 | 22.66 | 30.46 | 37.63 | 73.35 | <0.05 | <0.05 | <0.05 |
| + 3 Hours (50-60) | 29.05 | 10.16 | 12.56 | 21.42 | 27.85 | 34.11 | 65.09 | <0.05 | <0.05 | <0.05 |
| + 3 Hours (60-70) | 22.4 | 8.05 | 8.95 | 16.8 | 21.49 | 26.9 | 50.53 | <0.05 | <0.05 | <0.05 |
| + 6 Hours (40-50) | 19.15 | 7.77 | 1.39 | 13.61 | 18.95 | 23.2 | 51.98 | <0.05 | <0.05 | <0.05 |
| + 6 Hours (50-60) | 17.88 | 7.1 | 6.45 | 12.61 | 16.81 | 21.59 | 43.22 | <0.05 | <0.05 | <0.05 |
| + 6 Hours (60-70) | 14.16 | 5.86 | 4.84 | 9.74 | 13.06 | 17.47 | 37.68 | <0.05 | <0.05 | <0.05 |
| + 12 Hours (40-50) | 12.4 | 5.06 | 0.69 | 9.1 | 12.24 | 15.19 | 28.35 | <0.05 | <0.05 | <0.05 |
| + 12 Hours (50-60) | 11.53 | 4.91 | 3.24 | 8.28 | 10.33 | 14.46 | 31.28 | <0.05 | <0.05 | <0.05 |
| + 12 Hours (60-70) | 9.48 | 4.27 | 2.42 | 6.46 | 8.42 | 11.97 | 28.46 | <0.05 | <0.05 | <0.05 |
| + 24 Hours (40-50) | 9.24 | 4.27 | 0.35 | 6.16 | 8.98 | 12.01 | 23.81 | 0.059 | <0.05 | <0.05 |
| + 24 Hours (50-60) | 8.44 | 4 | 2.26 | 5.58 | 7.44 | 11.07 | 22.47 | <0.05 | <0.05 | <0.05 |
| + 24 Hours (60-70) | 7.25 | 3.73 | 1.7 | 4.7 | 6.1 | 9.6 | 21.46 | <0.05 | <0.05 | <0.05 |
| + 48 Hours (40-50) | 7.08 | 3.7 | 0.17 | 4.44 | 6.33 | 9.84 | 21.24 | <0.05 | <0.05 | <0.05 |
| + 48 Hours (50-60) | 6.68 | 3.72 | 1.25 | 4.22 | 5.79 | 8.91 | 20.63 | <0.05 | <0.05 | <0.05 |
| + 48 Hours (60-70) | 5.98 | 3.55 | 1.08 | 3.62 | 5.06 | 8.05 | 19.41 | <0.05 | <0.05 | <0.05 |
| - 48 Hours (40-54) | 6.4 | 3.89 | 0.86 | 3.56 | 5.71 | 8.95 | 20.96 | <0.05 | <0.05 | <0.05 |
| - 48 Hours (55-70) | 5.88 | 3.67 | 0.96 | 3.5 | 4.96 | 7.89 | 19.79 | <0.05 | <0.05 | <0.05 |
| - 24 Hours (40-54) | 7.91 | 4.25 | 1.43 | 4.69 | 6.9 | 10.45 | 23.26 | <0.05 | <0.05 | <0.05 |
| - 24 Hours (55-70) | 7.1 | 3.92 | 1.58 | 4.58 | 5.99 | 9.45 | 21.58 | <0.05 | <0.05 | <0.05 |
| - 12 Hours (40-54) | 10.43 | 5.22 | 1.19 | 6.76 | 9.74 | 13.11 | 32.4 | <0.05 | <0.05 | <0.05 |
| - 12 Hours (55-70) | 9.18 | 4.58 | 1.82 | 5.96 | 8.18 | 11.63 | 28.6 | <0.05 | <0.05 | <0.05 |
| - 6 Hours (40-54) | 15.89 | 7.78 | 2.38 | 10.11 | 14.93 | 19.93 | 45.74 | <0.05 | <0.05 | <0.05 |
| - 6 Hours (55-70) | 13.43 | 6.37 | 3.65 | 8.59 | 12.65 | 16.16 | 37.79 | <0.05 | <0.05 | <0.05 |
| - 3 Hours (40-54) | 25.55 | 11.9 | 4.76 | 16.1 | 24.08 | 31.77 | 61.8 | <0.05 | <0.05 | <0.05 |
| - 3 Hours (55-70) | 21.05 | 8.91 | 6.09 | 14.33 | 19.93 | 26.11 | 54.91 | <0.05 | <0.05 | <0.05 |
| + 3 Hours (40-54) | 30.99 | 11.9 | 10.34 | 22.29 | 30.19 | 38.08 | 72.83 | <0.05 | <0.05 | <0.05 |
| + 3 Hours (55-70) | 23.3 | 8.62 | 9.21 | 17.24 | 22.21 | 28.04 | 51.81 | <0.05 | <0.05 | <0.05 |
| + 6 Hours (40-54) | 18.74 | 8.01 | 6.21 | 12.74 | 18.34 | 22.83 | 50.41 | <0.05 | <0.05 | <0.05 |
| + 6 Hours (55-70) | 14.98 | 6.34 | 5.43 | 10.16 | 14.16 | 18.09 | 40.05 | <0.05 | <0.05 | <0.05 |
| + 12 Hours (40-54) | 12.2 | 5.29 | 3.57 | 8.49 | 11.99 | 14.87 | 33.24 | <0.05 | <0.05 | <0.05 |
| + 12 Hours (55-70) | 9.7 | 4.48 | 2.62 | 6.67 | 8.7 | 12.16 | 29.1 | <0.05 | <0.05 | <0.05 |
| + 24 Hours (40-54) | 8.93 | 4.31 | 1.81 | 5.68 | 8.71 | 11.71 | 23.87 | <0.05 | <0.05 | <0.05 |
| + 24 Hours (55-70) | 7.34 | 3.84 | 1.81 | 4.74 | 6.21 | 9.41 | 21.62 | <0.05 | <0.05 | <0.05 |
| + 48 Hours (40-54) | 6.93 | 3.88 | 1.36 | 4.13 | 6.01 | 9.64 | 21.5 | <0.05 | <0.05 | <0.05 |
| + 48 Hours (55-70) | 6.1 | 3.65 | 1.11 | 3.68 | 5.09 | 8.17 | 19.97 | <0.05 | <0.05 | <0.0 |

*The metrics include: mean, standard deviation (SD), minimum (Min), first quartile (25%), median (50%), third quartile (75%), maximum (Max), Shapiro-Wilk test (SW-test), Z-test, and Kolmogorov-Smirnov test (KS-test). The results of the SW-test, Z-test, and KS-test, provide information about the normality of the data and the equality of distribution, respectively. In all cases for SW-test except 24 hours ahead of hypoglycemia in range 40-50 mg/dL, $p<0.05$ indicates that the data does not follow a normal distribution. Furthermore, Z-test and KS-test indicate that the distributions differ significantly from each other ($p<0.05$).*

Table A2: **Statistics of glucose analysis and variability metrics for Time In Range (TIR) for various time periods surrounding the instances of hypoglycemia, respectively.**

| | Mean (%) | SD (%) | Min (%) | 25% | 50% | 75% | Max (%) | SW-test | Z-test | KS-test |
|---|---|---|---|---|---|---|---|---|---|---|
| Overall | 76.99 | 8.89 | 49.75 | 71.32 | 77.91 | 83.67 | 96.87 | 0.662 | | |
| - 48 Hours (40-50) | 74.89 | 9.25 | 38.68 | 70.64 | 75.2 | 80.84 | 93.91 | <0.05 | 0.098 | 0.293 |
| - 48 Hours (50-60) | 75.59 | 8.78 | 48.52 | 70.71 | 76.66 | 81.43 | 95.3 | 0.172 | 0.259 | 0.293 |
| - 48 Hours (60-70) | 75.59 | 8.78 | 48.52 | 70.71 | 76.66 | 81.43 | 95.3 | 0.172 | 0.259 | 0.293 |
| - 24 Hours (40-50) | 73.14 | 9.82 | 40.58 | 67.61 | 74.07 | 79.75 | 91.98 | <0.05 | <0.05 | 0.058 |
| - 24 Hours (50-60) | 74.59 | 8.45 | 49.9 | 69.41 | 75.27 | 80.01 | 94.78 | 0.321 | <0.05 | 0.118 |
| - 24 Hours (60-70) | 74.59 | 8.45 | 49.9 | 69.41 | 75.27 | 80.01 | 94.78 | 0.321 | <0.05 | 0.118 |
| - 12 Hours (40-50) | 71.27 | 10.1 | 38.8 | 67.07 | 71.39 | 77.48 | 90.14 | <0.05 | <0.05 | <0.05 |
| - 12 Hours (50-60) | 73.05 | 8.46 | 52.79 | 68.19 | 73.21 | 78.27 | 92.04 | 0.098 | <0.05 | <0.05 |
| - 12 Hours (60-70) | 75.32 | 8.22 | 50.33 | 70.45 | 75.73 | 80.67 | 93.81 | 0.085 | 0.164 | 0.163 |
| - 6 Hours (40-50) | 67.12 | 11.1 | 24.4 | 63.06 | 68.82 | 72.95 | 93.06 | <0.05 | <0.05 | <0.05 |
| - 6 Hours (50-60) | 69.63 | 8.53 | 47.41 | 65.66 | 70.3 | 74.52 | 87.36 | <0.05 | <0.05 | <0.05 |
| - 6 Hours (60-70) | 73.38 | 7.71 | 51.31 | 70.17 | 74.2 | 78.33 | 90.19 | <0.05 | <0.05 | <0.05 |
| - 3 Hours (40-50) | 62.25 | 11.1 | 28.17 | 56.92 | 63.53 | 70.25 | 86.11 | <0.05 | <0.05 | <0.05 |
| - 3 Hours (50-60) | 64.89 | 9.47 | 35.44 | 60.6 | 65.69 | 71.18 | 85.11 | <0.05 | <0.05 | <0.05 |
| - 3 Hours (60-70) | 64.89 | 9.47 | 35.44 | 60.6 | 65.69 | 71.18 | 85.11 | <0.05 | <0.05 | <0.05 |
| + 3 Hours (40-50) | 59.14 | 10.4 | 26.52 | 52.65 | 60.7 | 66.4 | 94.44 | <0.05 | <0.05 | <0.05 |
| + 3 Hours (50-60) | 63.26 | 8.71 | 34.74 | 58.85 | 64.44 | 68.3 | 78.98 | <0.05 | <0.05 | <0.05 |
| + 3 Hours (60-70) | 71.1 | 6.99 | 47.54 | 67.32 | 71.66 | 75.66 | 83.74 | <0.05 | <0.05 | <0.05 |
| + 6 Hours (40-50) | 64.81 | 9.98 | 39.93 | 58.95 | 65.08 | 70.45 | 97.22 | 0.22 | <0.05 | <0.05 |
| + 6 Hours (50-60) | 67.8 | 8.19 | 43.32 | 64.42 | 69.16 | 72.51 | 84.82 | <0.05 | <0.05 | <0.05 |
| + 6 Hours (60-70) | 72.94 | 6.97 | 56.41 | 69.31 | 73.28 | 77.42 | 87.17 | <0.05 | <0.05 | <0.05 |
| + 12 Hours (40-50) | 69.68 | 9.59 | 45.29 | 64.62 | 70.33 | 75.24 | 98.61 | 0.25 | <0.05 | <0.05 |
| + 12 Hours (50-60) | 71.64 | 8.31 | 47.63 | 66.85 | 72.36 | 76.8 | 91.04 | 0.355 | <0.05 | <0.05 |
| + 12 Hours (60-70) | 74.85 | 7.73 | 52.15 | 69.38 | 75.76 | 79.98 | 92.52 | 0.187 | 0.067 | <0.05 |
| + 24 Hours (40-50) | 73.02 | 8.96 | 51.39 | 67.97 | 73.45 | 77.5 | 99.31 | 0.335 | <0.05 | <0.05 |
| + 24 Hours (50-60) | 74.38 | 8.24 | 52.65 | 70.02 | 74.55 | 80.02 | 94.56 | 0.56 | <0.05 | 0.058 |
| + 24 Hours (60-70) | 76.38 | 7.84 | 55.7 | 72.11 | 77.26 | 81.8 | 94.96 | 0.268 | 0.604 | 0.38 |
| + 48 Hours (40-50) | 74.95 | 8.13 | 54.63 | 70.41 | 75.24 | 79.6 | 94.7 | 0.764 | 0.087 | 0.163 |
| + 48 Hours (50-60) | 75.78 | 8.07 | 53.19 | 71.14 | 76.11 | 81.11 | 94.8 | 0.389 | 0.311 | 0.38 |
| + 48 Hours (60-70) | 77.09 | 7.91 | 54.91 | 72.1 | 77.84 | 82.21 | 95.87 | 0.572 | 0.931 | 0.914 |
| - 48 Hours (40-54) | 75.16 | 9.45 | 39.04 | 69.56 | 75.87 | 81.39 | 94.38 | <0.05 | 0.159 | 0.298 |
| - 48 Hours (55-70) | 76.75 | 8.54 | 49.23 | 71.64 | 77.93 | 82.69 | 95.45 | 0.122 | 0.865 | 0.995 |
| - 24 Hours (40-54) | 73.91 | 9.41 | 44.09 | 68.73 | 74.29 | 80.76 | 93.03 | 0.169 | <0.05 | 0.086 |
| - 24 Hours (55-70) | 76.19 | 8.28 | 50.84 | 70.48 | 77.59 | 81.97 | 94.76 | 0.119 | 0.523 | 0.829 |
| - 12 Hours (40-54) | 72.23 | 9.83 | 40.93 | 66.14 | 72.22 | 78.03 | 95.14 | 0.718 | <0.05 | <0.05 |
| - 12 Hours (55-70) | 75.2 | 8.21 | 51.67 | 71.13 | 76.19 | 80.68 | 93.56 | 0.065 | 0.14 | 0.167 |
| - 6 Hours (40-54) | 68.46 | 10.3 | 32.1 | 63.14 | 68.98 | 74.23 | 91.32 | 0.223 | <0.05 | <0.05 |
| - 6 Hours (55-70) | 72.95 | 7.87 | 50.36 | 69.31 | 73.4 | 78.34 | 89.74 | <0.05 | <0.05 | <0.05 |
| - 3 Hours (40-54) | 63.74 | 11.1 | 35.46 | 58.02 | 63.74 | 71.34 | 90.47 | 0.07 | <0.05 | <0.05 |
| - 3 Hours (55-70) | 69.95 | 7.82 | 39.71 | 65.98 | 69.83 | 75.32 | 86.1 | <0.05 | <0.05 | <0.05 |
| + 3 Hours (40-54) | 60.03 | 10.2 | 26.99 | 53.26 | 61.35 | 66.99 | 86.51 | 0.134 | <0.05 | <0.05 |
| + 3 Hours (55-70) | 69.92 | 7.48 | 46.97 | 65.6 | 70.15 | 75.19 | 83.59 | <0.05 | <0.05 | <0.05 |
| + 6 Hours (40-54) | 65.56 | 9.33 | 44.99 | 59.53 | 66.49 | 71.55 | 86.7 | 0.169 | <0.05 | <0.05 |
| + 6 Hours (55-70) | 71.7 | 7.3 | 53.65 | 68.11 | 72.1 | 76.47 | 86.51 | 0.16 | <0.05 | <0.05 |
| + 12 Hours (40-54) | 70.53 | 9.01 | 44.94 | 65.46 | 71.21 | 76.8 | 87.6 | 0.158 | <0.05 | <0.05 |
| + 12 Hours (55-70) | 74.54 | 7.86 | 50.98 | 69.19 | 75.73 | 79.63 | 92.5 | 0.162 | <0.05 | <0.05 |
| + 24 Hours (40-54) | 73.55 | 8.69 | 50.21 | 67.61 | 73.83 | 79.34 | 93.12 | 0.852 | <0.05 | <0.05 |
| + 24 Hours (55-70) | 76.19 | 7.94 | 55.53 | 71.8 | 76.73 | 81.79 | 95.05 | 0.227 | 0.517 | 0.386 |
| + 48 Hours (40-54) | 75.27 | 8.35 | 54.08 | 69.75 | 75.7 | 80.88 | 94.35 | 0.873 | 0.16 | 0.225 |
| + 48 Hours (55-70) | 76.91 | 8.05 | 54.29 | 72.16 | 77.57 | 82.56 | 95.64 | 0.432 | 0.969 | 0.972 |

*The metrics include: mean, standard deviation (SD), minimum (Min), first quartile (25%), median (50%), third quartile (75%), maximum (Max), Shapiro-Wilk test (SW-test), Z-test, and Kolmogorov-Smirnov test (KS-test). The SW-test results indicate that the TIR distribution is not normally distributed for most of the groups ($p<0.05$). The Z-test results show significant differences ($p<0.05$) between the mean TIR of many groups and the overall mean TIR. The KS-test results show significant differences ($p<0.05$) in the TIR distribution between many groups and the overall TIR distribution.*

Table A3: **Statistics of glucose analysis and variability metrics for Time Above Range (TAR) or Time Outside Range greater than 180 mg/dL (TOR>180%) for various time periods surrounding the instances of hypoglycemia, respectively.**

| | Mean (%) | SD (%) | Min (%) | 25% | 50% | 75% | Max (%) | SW-test | Z-test | KS-test |
|---|---|---|---|---|---|---|---|---|---|---|
| Overall | 18.92 | 9.78 | 0.8 | 12.98 | 17.99 | 25.52 | 49.67 | 0.22 | | |
| - 48 Hours (40-50) | 18.63 | 10.15 | 0 | 11.24 | 18.36 | 23.77 | 53.05 | <0.05 | 0.836 | 0.971 |
| - 48 Hours (50-60) | 17.97 | 9.63 | 1.47 | 11.31 | 17.02 | 24.35 | 49.65 | 0.062 | 0.482 | 0.825 |
| - 48 Hours (60-70) | 17.97 | 9.63 | 1.47 | 11.31 | 17.02 | 24.35 | 49.65 | 0.062 | 0.482 | 0.825 |
| - 24 Hours (40-50) | 18.84 | 10.64 | 0 | 10.79 | 18.18 | 24.27 | 50.64 | <0.05 | 0.953 | 0.825 |
| - 24 Hours (50-60) | 17.42 | 9.28 | 1.09 | 10.92 | 15.99 | 22.92 | 47.62 | 0.061 | 0.261 | 0.596 |
| - 24 Hours (60-70) | 17.42 | 9.28 | 1.09 | 10.92 | 15.99 | 22.92 | 47.62 | 0.061 | 0.261 | 0.596 |
| - 12 Hours (40-50) | 18.15 | 10.59 | 0 | 9.98 | 17.6 | 23.89 | 55.28 | <0.05 | 0.588 | 0.596 |
| - 12 Hours (50-60) | 16.49 | 8.91 | 1.39 | 10.56 | 15.19 | 21.68 | 42.45 | <0.05 | 0.063 | 0.118 |
| - 12 Hours (60-70) | 15.61 | 9 | 1.12 | 9.74 | 14.1 | 20.52 | 45.76 | <0.05 | <0.05 | <0.05 |
| - 6 Hours (40-50) | 16.65 | 11.43 | 0 | 8.37 | 16 | 22.37 | 69.25 | <0.05 | 0.128 | 0.058 |
| - 6 Hours (50-60) | 14.64 | 8.84 | 1.14 | 8.52 | 14.6 | 18.69 | 46.01 | <0.05 | <0.05 | <0.05 |
| - 6 Hours (60-70) | 13.36 | 8.17 | 1.08 | 7.9 | 12.53 | 17.51 | 40.8 | <0.05 | <0.05 | <0.05 |
| - 3 Hours (40-50) | 11.83 | 9.96 | 0 | 5.4 | 9.94 | 15.22 | 59.26 | <0.05 | <0.05 | <0.05 |
| - 3 Hours (50-60) | 10.09 | 7.53 | 0.61 | 5.06 | 9.22 | 12.94 | 42.95 | <0.05 | <0.05 | <0.05 |
| - 3 Hours (60-70) | 10.09 | 7.53 | 0.61 | 5.06 | 9.22 | 12.94 | 42.95 | <0.05 | <0.05 | <0.05 |
| + 3 Hours (40-50) | 9.02 | 7.66 | 0 | 3.38 | 7.24 | 12.66 | 43.06 | <0.05 | <0.05 | <0.05 |
| + 3 Hours (50-60) | 7.69 | 6.04 | 0.16 | 3.72 | 6.28 | 10.2 | 37.73 | <0.05 | <0.05 | <0.05 |
| + 3 Hours (60-70) | 6.5 | 4.51 | 0.24 | 3.23 | 5.31 | 9.11 | 19.77 | <0.05 | <0.05 | <0.05 |
| + 6 Hours (40-50) | 16.04 | 10.15 | 0 | 8.4 | 14.74 | 22.91 | 52.48 | <0.05 | <0.05 | 0.083 |
| + 6 Hours (50-60) | 14.32 | 8.72 | 1.42 | 8.41 | 12.82 | 19.25 | 49.83 | <0.05 | <0.05 | <0.05 |
| + 6 Hours (60-70) | 12.9 | 7.78 | 0.85 | 7.16 | 11.19 | 17.92 | 33.2 | <0.05 | <0.05 | <0.05 |
| + 12 Hours (40-50) | 17.92 | 9.77 | 0.04 | 10.94 | 16.36 | 24.31 | 45.22 | 0.2 | 0.465 | 0.596 |
| + 12 Hours (50-60) | 16.83 | 8.94 | 0.77 | 11.05 | 15.54 | 22.25 | 43.52 | <0.05 | 0.111 | 0.163 |
| + 12 Hours (60-70) | 15.67 | 8.55 | 0.57 | 10.31 | 13.59 | 21.36 | 41.69 | <0.05 | <0.05 | <0.05 |
| + 24 Hours (40-50) | 17.74 | 9.26 | 0.14 | 11.54 | 17.15 | 23.79 | 42.44 | 0.295 | 0.374 | 0.714 |
| + 24 Hours (50-60) | 17.17 | 8.96 | 0.89 | 11.7 | 15.86 | 22.61 | 43.24 | 0.083 | 0.183 | 0.38 |
| + 24 Hours (60-70) | 16.37 | 8.7 | 0.42 | 11.09 | 14.74 | 21.28 | 41.94 | 0.116 | <0.05 | 0.058 |
| + 48 Hours (40-50) | 17.98 | 8.87 | 0.07 | 11.67 | 18.47 | 23.61 | 40.96 | 0.477 | 0.469 | 0.825 |
| + 48 Hours (50-60) | 17.53 | 8.92 | 0.79 | 11.54 | 17.03 | 22.6 | 44.52 | 0.192 | 0.289 | 0.38 |
| + 48 Hours (60-70) | 16.93 | 8.81 | 0.48 | 11.72 | 16.43 | 22.3 | 43.39 | 0.236 | 0.126 | 0.293 |
| - 48 Hours (40-54) | 18.44 | 10.4 | 0.13 | 10.88 | 17.68 | 24.51 | 51.61 | <0.05 | 0.707 | 0.829 |
| - 48 Hours (55-70) | 17.37 | 9.65 | 0.03 | 10.61 | 15.88 | 23.38 | 48.98 | 0.071 | 0.237 | 0.489 |
| - 24 Hours (40-54) | 18.19 | 10.17 | 0.26 | 9.77 | 17.34 | 23.65 | 49.54 | <0.05 | 0.572 | 0.719 |
| - 24 Hours (55-70) | 16.71 | 9.35 | 0.01 | 9.91 | 15.15 | 22.81 | 46.63 | 0.056 | 0.09 | 0.225 |
| - 12 Hours (40-54) | 17.34 | 10 | 0 | 9.83 | 17.21 | 23.21 | 49.33 | <0.05 | 0.236 | 0.225 |
| - 12 Hours (55-70) | 15.62 | 9.04 | 0.01 | 9.22 | 14.12 | 20.59 | 44.42 | <0.05 | <0.05 | <0.05 |
| - 6 Hours (40-54) | 15.65 | 10.43 | 0 | 8.11 | 14.84 | 20.49 | 62.52 | <0.05 | <0.05 | <0.05 |
| - 6 Hours (55-70) | 13.61 | 8.41 | 0 | 7.99 | 12.93 | 17.96 | 40.11 | <0.05 | <0.05 | <0.05 |
| - 3 Hours (40-54) | 10.72 | 8.95 | 0 | 4.83 | 9.21 | 12.6 | 53 | <0.05 | <0.05 | <0.05 |
| - 3 Hours (55-70) | 9 | 6.33 | 0.01 | 4.61 | 8.26 | 10.96 | 31.55 | <0.05 | <0.05 | <0.05 |
| + 3 Hours (40-54) | 8.97 | 7.64 | 0 | 3.3 | 6.98 | 13.29 | 36.67 | <0.05 | <0.05 | <0.05 |
| + 3 Hours (55-70) | 6.78 | 4.92 | 0.01 | 3.19 | 5.56 | 9.49 | 20.37 | <0.05 | <0.05 | <0.05 |
| + 6 Hours (40-54) | 15.7 | 9.59 | 0.59 | 8.1 | 14.28 | 22.64 | 48.06 | <0.05 | <0.05 | <0.05 |
| + 6 Hours (55-70) | 13.32 | 8.04 | 0 | 7.46 | 11.77 | 18.95 | 34.81 | <0.05 | <0.05 | <0.05 |
| + 12 Hours (40-54) | 17.27 | 9.31 | 0.66 | 10.79 | 16.08 | 22.81 | 44.89 | 0.148 | 0.2 | 0.489 |
| + 12 Hours (55-70) | 15.77 | 8.74 | 0.01 | 9.99 | 13.82 | 21.58 | 42.18 | 0.051 | <0.05 | <0.05 |
| + 24 Hours (40-54) | 17.52 | 9.43 | 0.26 | 10.78 | 16.77 | 24.32 | 43.78 | 0.161 | 0.28 | 0.602 |
| + 24 Hours (55-70) | 16.47 | 8.89 | 0.03 | 11.25 | 14.81 | 21.67 | 41.72 | 0.171 | 0.054 | 0.167 |
| + 48 Hours (40-54) | 17.81 | 9.31 | 0.17 | 11.35 | 17.85 | 24.48 | 42.65 | 0.369 | 0.382 | 0.602 |
| + 48 Hours (55-70) | 17 | 9.04 | 0.04 | 10.75 | 15.59 | 22.61 | 43.82 | 0.324 | 0.132 | 0.489 |

*The metrics include: mean, standard deviation (SD), minimum (Min), first quartile (25%), median (50%), third quartile (75%), maximum (Max), Shapiro-Wilk test (SW-test), Z-test, and Kolmogorov-Smirnov test (KS-test).*

Table A4: **Statistics of glucose analysis and variability metrics for Standard Deviation (SD) in mg/dL for various time periods surrounding the instances of hypoglycemia, respectively.**

| | Mean (%) | SD (%) | Min (%) | 25% | 50% | 75% | Max (%) | SW-test | Z-test | KS-test |
|---|---|---|---|---|---|---|---|---|---|---|
| Overall | 50.16 | 10.8 | 21.47 | 43.22 | 50.23 | 58.47 | 77.33 | 0.831 | | |
| - 48 Hours (40-50) | 48.04 | 10.39 | 26.31 | 40.74 | 47.36 | 54.26 | 76.52 | 0.53 | 0.152 | 0.293 |
| - 48 Hours (50-60) | 47.03 | 10.29 | 23.23 | 40.32 | 46.67 | 54.06 | 74.72 | 0.91 | <0.05 | 0.083 |
| - 48 Hours (60-70) | 47.03 | 10.29 | 23.23 | 40.32 | 46.67 | 54.06 | 74.72 | 0.91 | <0.05 | 0.083 |
| - 24 Hours (40-50) | 47.07 | 10.27 | 26.79 | 40.28 | 46.08 | 53.88 | 76.52 | 0.113 | <0.05 | <0.05 |
| - 24 Hours (50-60) | 45.3 | 9.77 | 23.04 | 39.25 | 44.41 | 52.21 | 71.14 | 0.903 | <0.05 | <0.05 |
| - 24 Hours (60-70) | 45.3 | 9.77 | 23.04 | 39.25 | 44.41 | 52.21 | 71.14 | 0.903 | <0.05 | <0.05 |
| - 12 Hours (40-50) | 44.5 | 10.53 | 26.13 | 37.82 | 43.92 | 50.5 | 84.25 | <0.05 | <0.05 | <0.05 |
| - 12 Hours (50-60) | 42.18 | 9.49 | 22.13 | 35.63 | 41.85 | 47.81 | 70.43 | 0.855 | <0.05 | <0.05 |
| - 12 Hours (60-70) | 40.32 | 9.57 | 20.57 | 33.98 | 40.49 | 47.14 | 69.7 | 0.61 | <0.05 | <0.05 |
| - 6 Hours (40-50) | 39.53 | 9.86 | 12.36 | 33.02 | 39.17 | 44.45 | 80.8 | <0.05 | <0.05 | <0.05 |
| - 6 Hours (50-60) | 37.29 | 8.45 | 20.98 | 32.05 | 37.45 | 42.26 | 58.62 | 0.299 | <0.05 | <0.05 |
| - 6 Hours (60-70) | 34.89 | 8.64 | 17.95 | 29.52 | 34.31 | 40.9 | 57.46 | 0.414 | <0.05 | <0.05 |
| - 3 Hours (40-50) | 34.43 | 9.48 | 7.72 | 28.26 | 33.88 | 39.06 | 73.79 | <0.05 | <0.05 | <0.05 |
| - 3 Hours (50-60) | 31.39 | 7.82 | 14.33 | 25.57 | 30.45 | 36.02 | 55.23 | <0.05 | <0.05 | <0.05 |
| - 3 Hours (60-70) | 31.39 | 7.82 | 14.33 | 25.57 | 30.45 | 36.02 | 55.23 | <0.05 | <0.05 | <0.05 |
| + 3 Hours (40-50) | 34.69 | 9.41 | 11.73 | 29.56 | 33.2 | 41.08 | 65.49 | 0.19 | <0.05 | <0.05 |
| + 3 Hours (50-60) | 30.91 | 8.14 | 11.72 | 26.26 | 30.07 | 36.08 | 57.16 | 0.275 | <0.05 | <0.05 |
| + 3 Hours (60-70) | 26.95 | 7.15 | 11.61 | 22.2 | 26.99 | 32.2 | 45.09 | 0.773 | <0.05 | <0.05 |
| + 6 Hours (40-50) | 43.23 | 10.47 | 21.69 | 35.84 | 43.14 | 49.39 | 64.35 | 0.168 | <0.05 | <0.05 |
| + 6 Hours (50-60) | 39.66 | 9.35 | 19.78 | 34.05 | 39.21 | 44.97 | 68.05 | 0.331 | <0.05 | <0.05 |
| + 6 Hours (60-70) | 36.07 | 8.95 | 17.97 | 30.2 | 35.23 | 42.02 | 58.45 | 0.617 | <0.05 | <0.05 |
| + 12 Hours (40-50) | 46.3 | 10.57 | 18.14 | 39.27 | 46.25 | 54.5 | 69.86 | 0.734 | <0.05 | <0.05 |
| + 12 Hours (50-60) | 43.95 | 9.54 | 20.15 | 37.89 | 43.98 | 50.05 | 71.75 | 0.992 | <0.05 | <0.05 |
| + 12 Hours (60-70) | 41.32 | 9.38 | 17.83 | 35.67 | 41.41 | 46.89 | 67.7 | 0.948 | <0.05 | <0.05 |
| + 24 Hours (40-50) | 47.4 | 10.51 | 14.69 | 40.72 | 46.83 | 54.8 | 69.63 | 0.387 | 0.064 | 0.293 |
| + 24 Hours (50-60) | 45.83 | 9.63 | 21.31 | 38.97 | 45.55 | 52.53 | 70.2 | 0.991 | <0.05 | <0.05 |
| + 24 Hours (60-70) | 44.07 | 9.55 | 19.08 | 38.34 | 43.7 | 50.58 | 67.51 | 0.929 | <0.05 | <0.05 |
| + 48 Hours (40-50) | 48.03 | 9.96 | 19.88 | 41.93 | 47.63 | 56.16 | 69.39 | 0.372 | 0.142 | 0.596 |
| + 48 Hours (50-60) | 47.04 | 9.7 | 20.5 | 40.72 | 46.97 | 53.55 | 69.03 | 0.841 | <0.05 | 0.083 |
| + 48 Hours (60-70) | 45.83 | 9.71 | 19.89 | 40.49 | 45.44 | 52.69 | 68.33 | 0.709 | <0.05 | <0.05 |
| - 48 Hours (40-54) | 47.43 | 10.99 | 16.16 | 40.63 | 46.81 | 55.05 | 76.87 | 0.978 | 0.064 | 0.167 |
| - 48 Hours (55-70) | 45.73 | 10.76 | 12.81 | 38.74 | 44.78 | 52.43 | 75.34 | 0.924 | <0.05 | <0.05 |
| - 24 Hours (40-54) | 46.1 | 10.62 | 17.35 | 39.4 | 44.5 | 52.37 | 71.85 | 0.582 | <0.05 | <0.05 |
| - 24 Hours (55-70) | 43.77 | 10.35 | 12.57 | 37.24 | 42.44 | 50.52 | 68.49 | 0.879 | <0.05 | <0.05 |
| - 12 Hours (40-54) | 43.26 | 10.5 | 21.34 | 35.88 | 42.71 | 49.85 | 77.83 | 0.296 | <0.05 | <0.05 |
| - 12 Hours (55-70) | 40.2 | 10.02 | 11.73 | 34.06 | 38.89 | 46.4 | 65.97 | 0.718 | <0.05 | <0.05 |
| - 6 Hours (40-54) | 38.64 | 9.45 | 19.16 | 31.9 | 38.55 | 45.09 | 67.15 | 0.233 | <0.05 | <0.05 |
| - 6 Hours (55-70) | 35.07 | 8.96 | 10.83 | 29.74 | 34.66 | 40.54 | 56.56 | 0.475 | <0.05 | <0.05 |
| - 3 Hours (40-54) | 33.33 | 8.71 | 16.34 | 27.71 | 32.33 | 37.18 | 69.22 | <0.05 | <0.05 | <0.05 |
| - 3 Hours (55-70) | 28.78 | 7.8 | 8.31 | 23.5 | 28.65 | 32.07 | 50.74 | 0.518 | <0.05 | <0.05 |
| + 3 Hours (40-54) | 34.25 | 9.59 | 11.74 | 28.67 | 32.42 | 40.32 | 60.83 | 0.384 | <0.05 | <0.05 |
| + 3 Hours (55-70) | 27.8 | 7.93 | 6.86 | 22.57 | 27.3 | 33.57 | 46.93 | 0.831 | <0.05 | <0.05 |
| + 6 Hours (40-54) | 42.39 | 10.37 | 20.95 | 35.31 | 41.09 | 48.61 | 65.66 | 0.202 | <0.05 | <0.05 |
| + 6 Hours (55-70) | 37 | 9.5 | 10.16 | 31.09 | 36.06 | 43.4 | 61.53 | 0.964 | <0.05 | <0.05 |
| + 12 Hours (40-54) | 45.39 | 10.26 | 22.13 | 38.14 | 45.18 | 53.19 | 70.76 | 0.684 | <0.05 | <0.05 |
| + 12 Hours (55-70) | 41.55 | 9.97 | 12.18 | 35.29 | 41.63 | 47.13 | 68.9 | 0.95 | <0.05 | <0.05 |
| + 24 Hours (40-54) | 46.82 | 10.57 | 17.24 | 39.53 | 46.53 | 54.39 | 70.29 | 0.882 | <0.05 | 0.086 |
| + 24 Hours (55-70) | 44.11 | 10.24 | 12.85 | 37.73 | 43.35 | 50.82 | 68.02 | 0.844 | <0.05 | <0.05 |
| + 48 Hours (40-54) | 47.51 | 10.47 | 17.19 | 40.46 | 47.12 | 55.32 | 69.61 | 0.489 | 0.065 | 0.167 |
| + 48 Hours (55-70) | 45.81 | 10.38 | 13.42 | 40.07 | 45.69 | 52.96 | 67.79 | 0.605 | <0.05 | <0.05 |

*The metrics include: mean, standard deviation (SD), minimum (Min), first quartile (25%), median (50%), third quartile (75%), maximum (Max), Shapiro-Wilk test (SW-test), Z-test, and Kolmogorov-Smirnov test (KS-test). The SD is smallest at the time of the hypoglycemia, as this is when glucose levels reach their lowest. Following the hypoglycemia, the SD begins to increase again, indicating that glucose levels become more variable. For example, the mean SD for the 40-50 mg/dL category increases from 34.69 mg/dL at +3 hours to 48.03 mg/dL at +48 hours. More severe hypoglycemia (lower glucose levels) are associated with smaller SD values both before and after the event. This suggests that more severe hypoglycemia is associated with less variable glucose levels. For SD, the data does not follow a normal distribution and have a statistically significant difference (generally p<0.05) in distribution when compared with normal (or overall) distribution.*

Table A5: **Statistics of glucose analysis and variability metrics for J_Index in mg/dL for various time periods surrounding the instances of hypoglycemia, respectively.**

| | Mean (%) | SD (%) | Min (%) | 25% | 50% | 75% | Max (%) | SW-test | Z-test | KS-test |
|---|---|---|---|---|---|---|---|---|---|---|
| Overall | 36.48 | 10.62 | 15.23 | 29.87 | 35.49 | 43.84 | 73.93 | 0.185 | | |
| - 48 Hours (40-50) | 35.76 | 11 | 15.88 | 27.95 | 35.6 | 41.15 | 74.26 | <0.05 | 0.631 | 0.596 |
| - 48 Hours (50-60) | 34.89 | 10.45 | 15.67 | 27.83 | 33.87 | 41.23 | 73.45 | <0.05 | 0.28 | 0.293 |
| - 48 Hours (60-70) | 34.89 | 10.45 | 15.67 | 27.83 | 33.87 | 41.23 | 73.45 | <0.05 | 0.28 | 0.293 |
| - 24 Hours (40-50) | 35.43 | 11.16 | 15.8 | 27.19 | 34.45 | 40.08 | 69.88 | <0.05 | 0.49 | 0.293 |
| - 24 Hours (50-60) | 33.83 | 9.98 | 15.1 | 27.27 | 32.39 | 39.91 | 70.4 | <0.05 | 0.066 | 0.118 |
| - 24 Hours (60-70) | 33.83 | 9.98 | 15.1 | 27.27 | 32.39 | 39.91 | 70.4 | <0.05 | 0.066 | 0.118 |
| - 12 Hours (40-50) | 34.2 | 12.01 | 14.66 | 25.71 | 33.07 | 39.5 | 87.75 | <0.05 | 0.151 | 0.163 |
| - 12 Hours (50-60) | 32.16 | 9.77 | 15 | 25.88 | 30.55 | 38.01 | 61.53 | <0.05 | <0.05 | <0.05 |
| - 12 Hours (60-70) | 31.32 | 10.12 | 14.33 | 25.39 | 29.45 | 36.05 | 69.37 | <0.05 | <0.05 | <0.05 |
| - 6 Hours (40-50) | 30.96 | 12.47 | 6.81 | 22.99 | 30.18 | 34.8 | 98.55 | <0.05 | <0.05 | <0.05 |
| - 6 Hours (50-60) | 28.86 | 9.42 | 13.21 | 22.65 | 28.15 | 33.72 | 59.69 | <0.05 | <0.05 | <0.05 |
| - 6 Hours (60-70) | 27.69 | 9.13 | 12.53 | 21.93 | 26.92 | 31.2 | 57.25 | <0.05 | <0.05 | <0.05 |
| - 3 Hours (40-50) | 24.92 | 10.19 | 5.27 | 18.74 | 23.29 | 28.83 | 72.29 | <0.05 | <0.05 | <0.05 |
| - 3 Hours (50-60) | 22.94 | 8.34 | 9.2 | 17.23 | 21.84 | 26.14 | 55.78 | <0.05 | <0.05 | <0.05 |
| - 3 Hours (60-70) | 22.94 | 8.34 | 9.2 | 17.23 | 21.84 | 26.14 | 55.78 | <0.05 | <0.05 | <0.05 |
| + 3 Hours (40-50) | 21.74 | 8.07 | 5.95 | 16.68 | 20.11 | 27.2 | 53.37 | <0.05 | <0.05 | <0.05 |
| + 3 Hours (50-60) | 20.27 | 6.77 | 7.22 | 16.21 | 19.4 | 24.61 | 48.49 | <0.05 | <0.05 | <0.05 |
| + 3 Hours (60-70) | 19.08 | 5.57 | 8.7 | 15.36 | 18.54 | 23.1 | 35.01 | 0.288 | <0.05 | <0.05 |
| + 6 Hours (40-50) | 30.82 | 10.64 | 12.08 | 23.61 | 29.59 | 37.14 | 70.51 | <0.05 | <0.05 | <0.05 |
| + 6 Hours (50-60) | 28.72 | 9.26 | 12.8 | 22.73 | 27.48 | 33.47 | 64.19 | <0.05 | <0.05 | <0.05 |
| + 6 Hours (60-70) | 27.02 | 8.32 | 12.03 | 21.41 | 25.51 | 32.72 | 50.88 | 0.07 | <0.05 | <0.05 |
| + 12 Hours (40-50) | 33.91 | 10.36 | 13.13 | 26.86 | 33.03 | 40.56 | 70.34 | 0.097 | 0.079 | <0.05 |
| + 12 Hours (50-60) | 32.55 | 9.52 | 14.37 | 26.52 | 31.29 | 38.36 | 62.32 | <0.05 | <0.05 | <0.05 |
| + 12 Hours (60-70) | 31.1 | 9.01 | 13.77 | 25.68 | 29.42 | 37 | 60.21 | 0.089 | <0.05 | <0.05 |
| + 24 Hours (40-50) | 34.37 | 9.95 | 13.9 | 27.47 | 33.72 | 40.98 | 64.7 | 0.354 | 0.143 | 0.221 |
| + 24 Hours (50-60) | 33.61 | 9.43 | 15.1 | 27.28 | 33.21 | 38.71 | 59.59 | 0.107 | <0.05 | 0.083 |
| + 24 Hours (60-70) | 32.67 | 9.14 | 14.53 | 26.83 | 31.34 | 38.3 | 58.26 | 0.267 | <0.05 | <0.05 |
| + 48 Hours (40-50) | 34.92 | 9.43 | 14.34 | 28.26 | 34.67 | 41.28 | 59.33 | 0.887 | 0.265 | 0.482 |
| + 48 Hours (50-60) | 34.38 | 9.41 | 14.74 | 27.91 | 34.46 | 39.83 | 60.89 | 0.568 | 0.135 | 0.221 |
| + 48 Hours (60-70) | 33.71 | 9.31 | 14.87 | 27.79 | 32.87 | 39.78 | 61.72 | 0.473 | <0.05 | 0.083 |
| - 48 Hours (40-54) | 35.42 | 11.26 | 11.12 | 27.59 | 34.66 | 41.81 | 73.37 | <0.05 | 0.484 | 0.719 |
| - 48 Hours (55-70) | 34.17 | 10.53 | 9.71 | 27.34 | 32.8 | 40.72 | 73.18 | 0.121 | 0.116 | 0.298 |
| - 24 Hours (40-54) | 34.64 | 10.83 | 11.88 | 26.42 | 33.75 | 40.61 | 67.7 | <0.05 | 0.215 | 0.298 |
| - 24 Hours (55-70) | 33.06 | 10.13 | 9.6 | 26.35 | 31.49 | 38.53 | 66.99 | 0.212 | <0.05 | <0.05 |
| - 12 Hours (40-54) | 33.25 | 11.01 | 13.82 | 24.77 | 32.6 | 40.01 | 75.92 | <0.05 | <0.05 | 0.086 |
| - 12 Hours (55-70) | 31.22 | 10.13 | 9.22 | 25.19 | 29.32 | 35.51 | 62.78 | <0.05 | <0.05 | <0.05 |
| - 6 Hours (40-54) | 30.01 | 10.77 | 11.96 | 22.53 | 29.31 | 34.73 | 78.49 | <0.05 | <0.05 | <0.05 |
| - 6 Hours (55-70) | 27.85 | 9.35 | 8.61 | 22.09 | 27.16 | 31.69 | 58.47 | <0.05 | <0.05 | <0.05 |
| - 3 Hours (40-54) | 23.95 | 9.55 | 9.22 | 18.31 | 22.54 | 26.64 | 71.4 | <0.05 | <0.05 | <0.05 |
| - 3 Hours (55-70) | 21.81 | 7.33 | 7.22 | 16.68 | 21.83 | 24.59 | 48.66 | <0.05 | <0.05 | <0.05 |
| + 3 Hours (40-54) | 21.73 | 8.35 | 6.52 | 16.48 | 19.84 | 27.71 | 50.17 | <0.05 | <0.05 | <0.05 |
| + 3 Hours (55-70) | 19.47 | 6.13 | 6.62 | 15.64 | 18.82 | 23.45 | 34.66 | 0.157 | <0.05 | <0.05 |
| + 6 Hours (40-54) | 30.3 | 10.27 | 12.33 | 22.87 | 28.81 | 36.88 | 62.8 | <0.05 | <0.05 | <0.05 |
| + 6 Hours (55-70) | 27.56 | 8.76 | 8.17 | 21.58 | 26.61 | 33.64 | 50.99 | 0.257 | <0.05 | <0.05 |
| + 12 Hours (40-54) | 33.17 | 10 | 13 | 26.9 | 31.96 | 40.43 | 61.35 | 0.201 | <0.05 | <0.05 |
| + 12 Hours (55-70) | 31.24 | 9.35 | 9.21 | 24.6 | 29.76 | 37.15 | 60.81 | 0.351 | <0.05 | <0.05 |
| + 24 Hours (40-54) | 34.08 | 10.12 | 10.87 | 27.66 | 33.45 | 40.57 | 60.79 | 0.373 | 0.095 | 0.167 |
| + 24 Hours (55-70) | 32.72 | 9.52 | 9.65 | 26.97 | 31.57 | 38.57 | 58.34 | 0.62 | <0.05 | <0.05 |
| + 48 Hours (40-54) | 34.7 | 9.94 | 11.21 | 28.1 | 33.81 | 41.28 | 59.99 | 0.955 | 0.212 | 0.602 |
| + 48 Hours (55-70) | 33.75 | 9.72 | 12.32 | 27.36 | 32.75 | 40.06 | 61.47 | 0.825 | 0.053 | 0.167 |

*The metrics include: mean, standard deviation (SD), minimum (Min), first quartile (25%), median (50%), third quartile (75%), maximum (Max), Shapiro-Wilk test (SW-test), Z-test, and Kolmogorov-Smirnov test (KS-test). Overall, for the J_index the mean score is 36.48 mg/dL, with a standard deviation (SD) of 10.62 mg/dL. The minimum score is 15.23 mg/dL, and the maximum score is 73.93 mg/dL. There is a non-normal (p<0.05) distribution of scores for most time periods and levels of hypoglycemia and most show a significant (p<0.005) difference from the overall distribution.*

**Figure A2:** Comparison of distributions (box plots) for A) LBGI and B) HBGI before and after hypoglycemia at levels of 40-50 mg/dL, 50-60 mg/dL, and 60-70 mg/dL.

*A) Distributions for LBGI before and after hypoglycemia.*

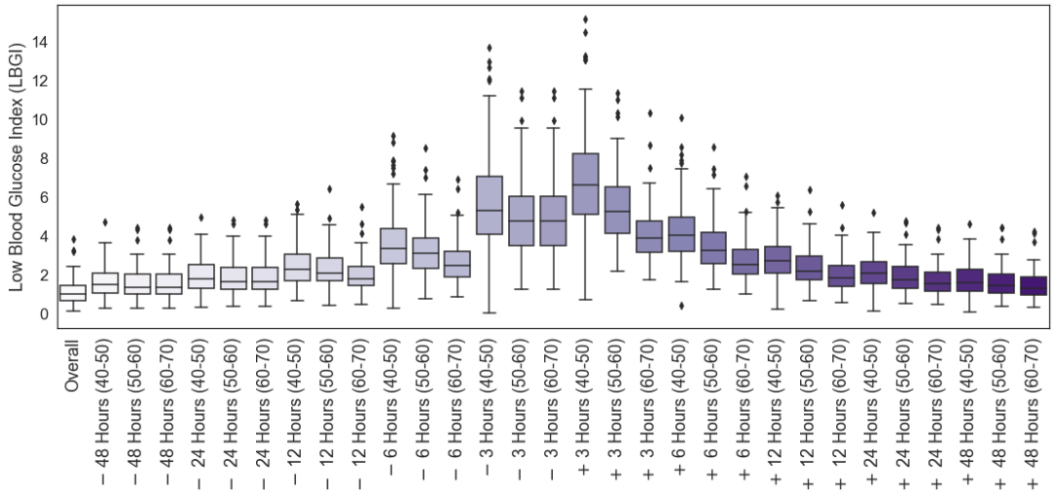

*B) Distributions for HBGI before and after hypoglycemia.*

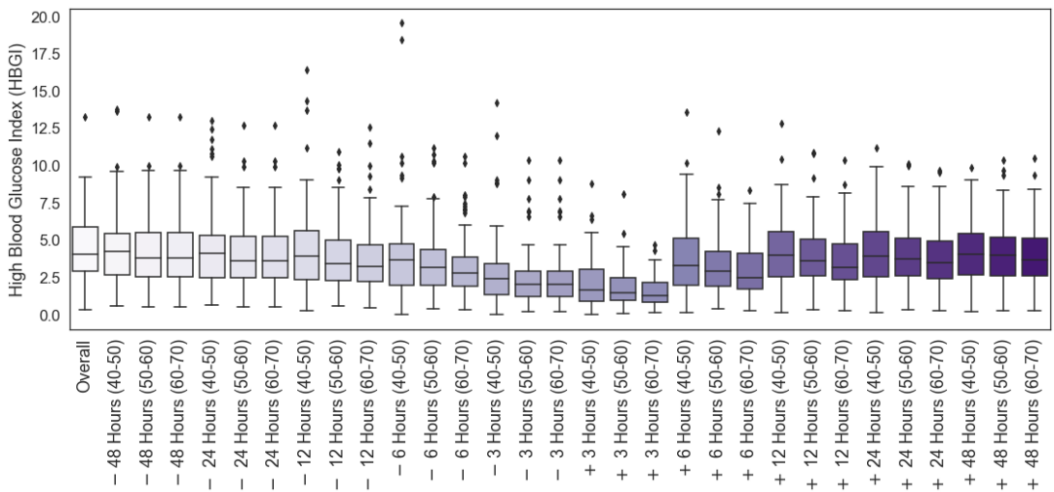

**Figure A3:** Comparison of distributions (box plots) for percentage of A) time below range (TBR or TOR<70), B) time in range (TIR), and C) time above range (TAR), before and after hypoglycemia at levels of 40-54 mg/dL and 55-70 mg/dL.

A) *Distributions for percentage of Time Below Range (TBR or TOR<70) before and after hypoglycemia.*

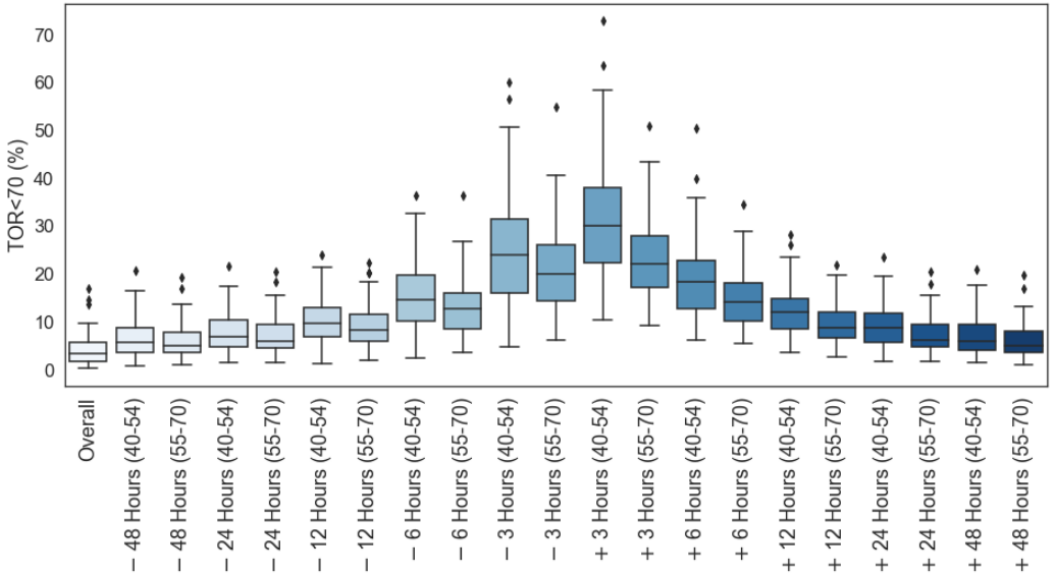

B) *Distributions for percentage of time in range (TIR) before and after hypoglycemia.*

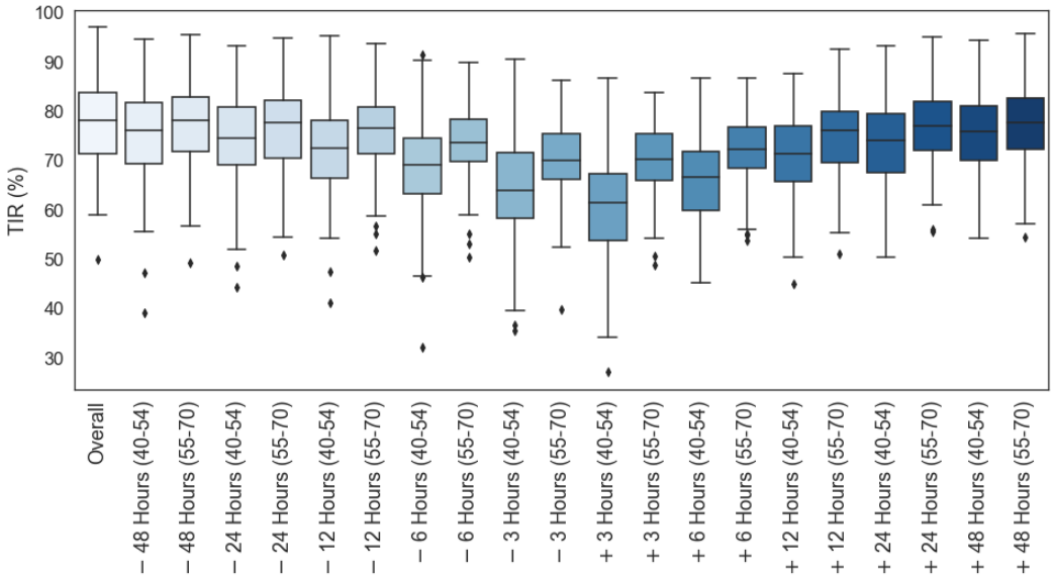

C) *Distributions for percentage of time above range (TAR or TOR>180) before and after hypoglycemia.*

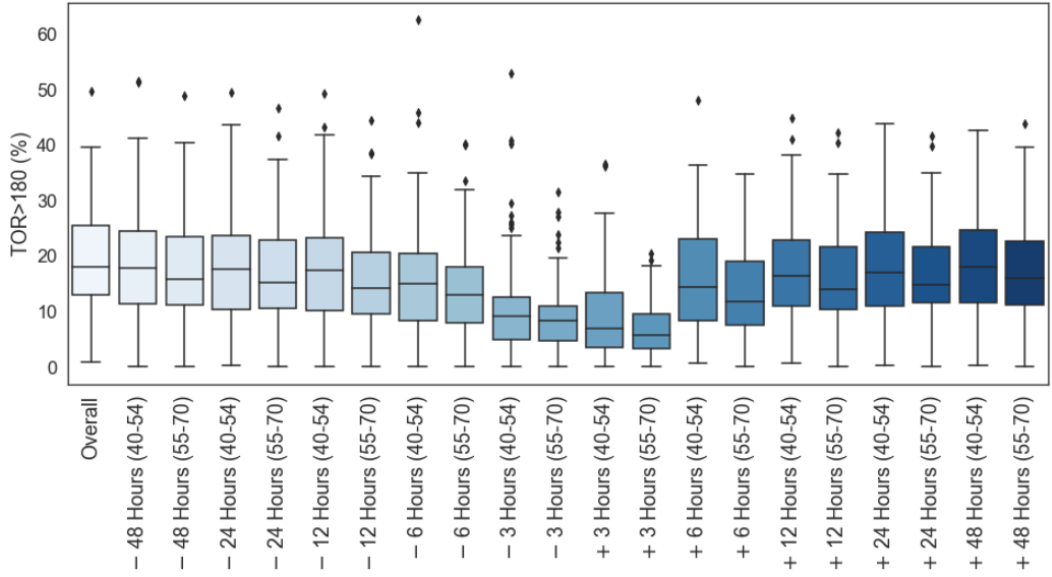

**Figure A4:** Comparison of distributions (box plots) for A) standard deviation
(SD) and B) J_Index, before and after hypoglycemia at levels of 40-54 mg/dL and
55-70 mg/dL

*A)  Distributions for standard deviation (SD) before and after hypoglycemia.*

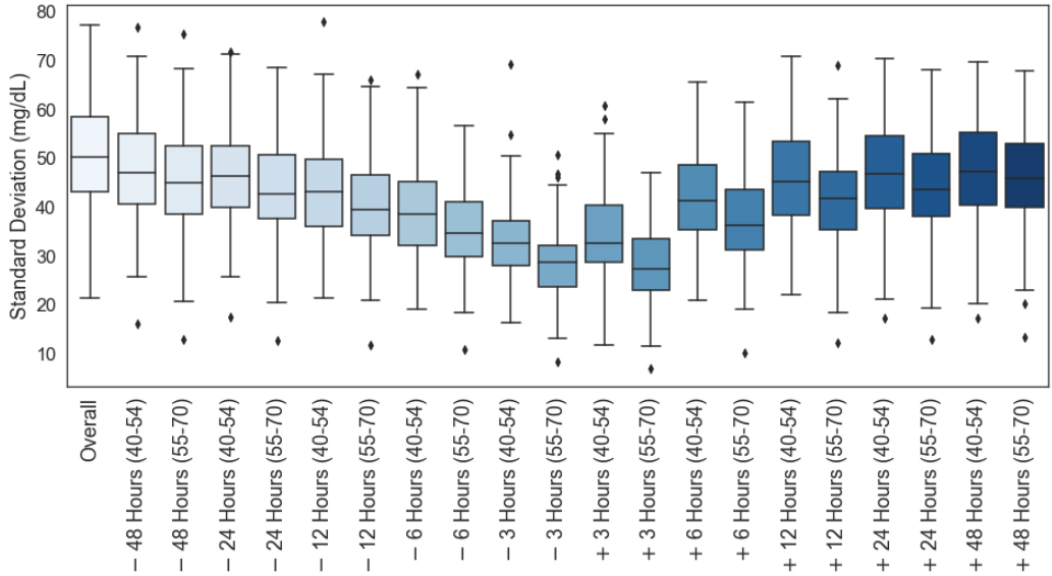

*B)  Distributions for J_Index before and after hypoglycemia.*

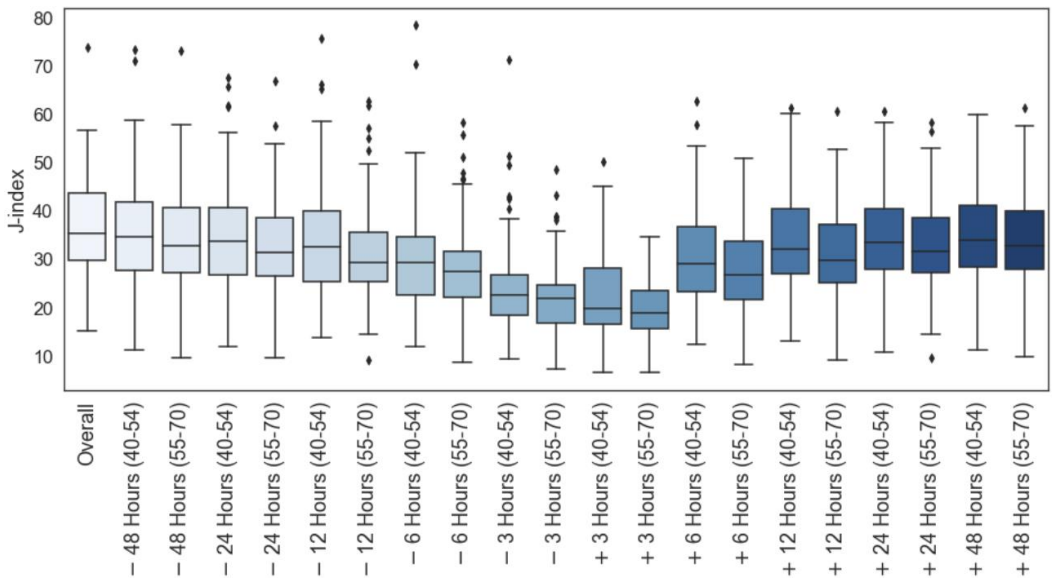

**Figure A5:** Comparison of distributions (box plots) for A) LBGI and B) HBGI before and after hypoglycemia at levels of 40-54 mg/dL and 55-70 mg/dL.

A)  *Distributions for LBGI before and after hypoglycemia*

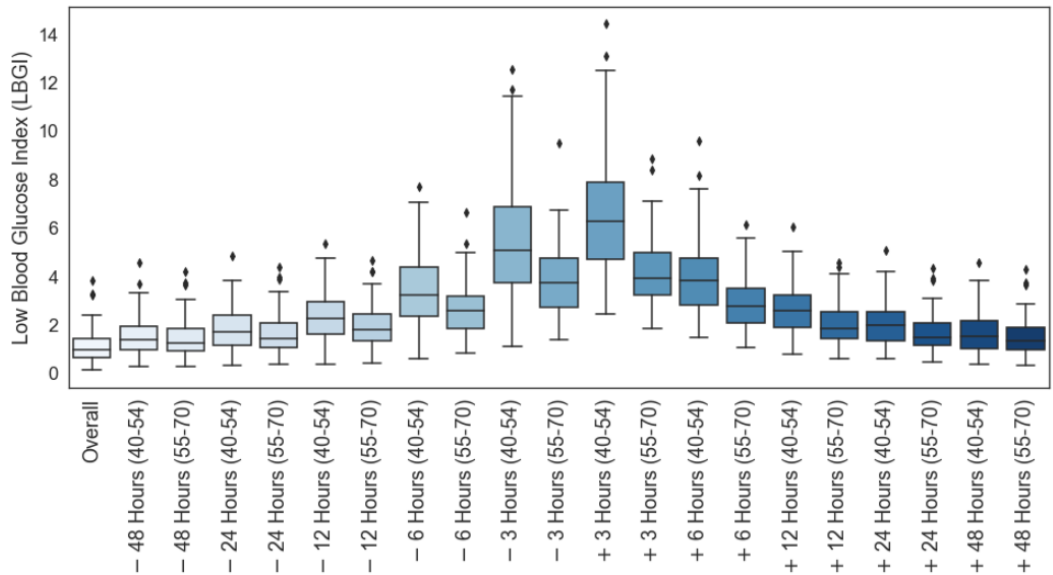

B)  *Distributions for HBGI before and after hypoglycemia.*

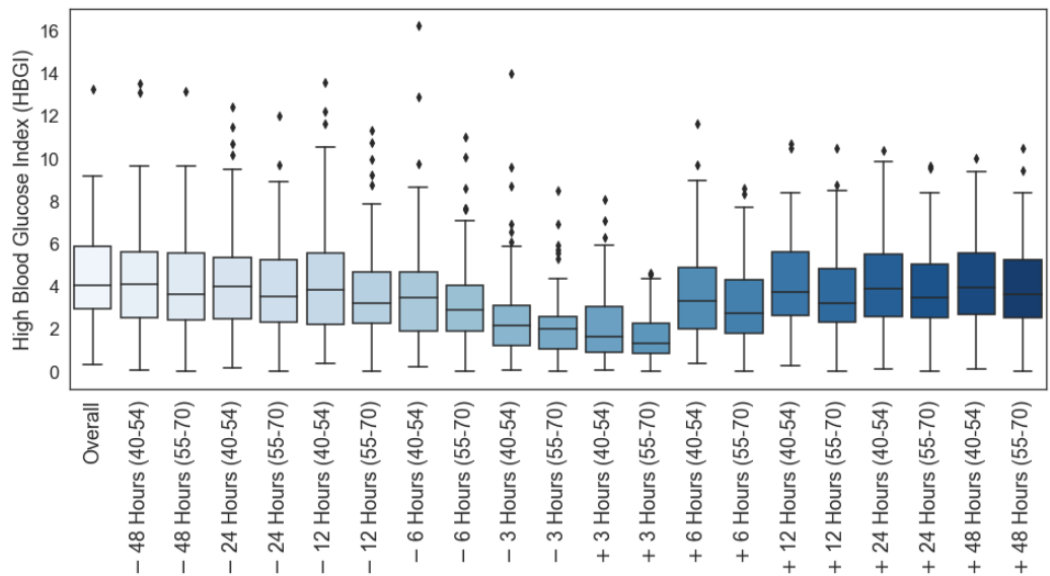